\documentclass[epj]{svjour}
\usepackage{graphics}
\usepackage{epsfig,array}
\usepackage{color}
\newcommand{\be}{\begin{eqnarray}}
\newcommand{\ee}{\end{eqnarray}}

\begin{document}
\title{Study of the proton-proton collisions at 1683 MeV/c}

\author{K.~N.~Ermakov      \and
        V.~A.~Nikonov      \and
        O.~V.~Rogachevsky  \and
        A.~V.~Sarantsev    \and
        V.~V.~Sarantsev   \mail{saran@pnpi.spb.ru} \and
        S.~G.~Sherman
}

\institute{Petersburg Nuclear Physics Institute NRC KI, Gatchina
188300, Russia }

\abstract{The new data on the elastic $pp$ and single pion
production reaction $pp\to pn \pi^{+}$ taken at the incident proton
momentum 1683 MeV/c are presented. The data on the $pp\to pn\pi^{+}$
reaction are compared with predictions from the OPE model. To extract
contributions of the leading partial waves the single pion production
data are analyzed in the framework of the event-by-event maximum
likelihood method together with the data measured earlier. }

\authorrunning{K.~N.~Ermakov {\it et al.} }
\titlerunning{Study of the proton-proton collisions at 1683 MeV/c}
\date{Received: date / Revised version: date}

 \maketitle

\section{Introduction}
\label{intro}
Understanding the proton-proton interaction at low and intermediate
energies is the important task of the particle physics. At large
momentum transfers where the strong coupling is small, the QCD
calculations can be used efficiently for the description of such
processes. A large prog\-ress was made at low energies where the
effective field approach allowed us to describe processes below the
resonance region. However the region of the intermediate energies
and especially the resonance region is much less accessible for the
theoretical calculations and phenomenological dynamic models play
the leading role  here. The data from the $NN$ collision reactions
forms the basis for the construction of such models which, in turn,
have the large range of applications in the nuclear and heavy ion
physics.

  In the region above the two pion production threshold and up to 1 GeV
the $NN\to\pi NN$ reaction is dominated by the production of the
$\Delta(1232)$ isobar in the intermediate state. It was natural to
suggest that such production is based on the one pion exchange
mechanism (OPE) and a set of the corresponding models was put
forward \cite{Selleri,Suslenko:1984qp,Dmitriev} a rather long time
ago. The pion exchange amplitudes are introduced there using certain
form factors with parameters defined from the fit of experimental
data. The model of Suslenko et al. describes with a reasonable
accuracy (up to normalization factors) the differential spectra of
the $pp\to pn\pi^+$ and $pp\to pp\pi^0$ reactions in the energy
rigion below 1 GeV \cite{Suslenko:1984qp,Andreev:1988fj}, while the
model of Dmitriev et al. was applied to the energies over 1 GeV
\cite{Dmitriev}. In the more complicated model based on the one
boson exchange mechanism \cite{Engel} the dominant contribution for
the $\Delta(1232)$ production is also defined by the one pion
exchange: it was found that other boson exchanges contribute around
10\% to the total cross section at the energies above 1 GeV. However
it should be noted that there are discrepancies in the simultaneous
description of the measured total cross sections for the $pp\to
pn\pi^+$ and $pp\to pp\pi^0$ reactions by the OPE model. For
example, at the proton momentum 1683 MeV/c, the OPE model can
reproduce well the $pp\to pn\pi^+$ measured total cross section with
the corresponding choice of the form factor. However, in this case
the OPE prediction for the $pp\to pp\pi^0$ total cross section will
be smaller by about 30\% than the experimental one (see
Ref.~\cite{Andreev:1988fj}).

Moreover in the region above the incident proton momentum 1.5 GeV/c
other contributions start to play a notable role: for example the
relatively broad Roper resonance is traced in the spectrum.
Therefore for a comprehensive analysis of data it is necessary to
apply an approach beyond the OPE model.

With this purpose we perform the partial wave analysis of the data
on the single pion production in the framework of the approach based
on the work \cite{Anisovich:2007zz}. The result of such an analysis
for the lower energy data measured earlier was reported in
\cite{Ermakov:2011at},\cite{Ermakov:2014aj}.

In the analysis \cite{Ermakov:2014aj} the several solutions had been
found which almost equally described the data. These solutions
differ by contributions from the partial waves with high orbital
momenta $L>3$ which were found to be rather unstable in the fit. It
is interesting to compare which solution is compatible with the
present higher energy data. It is also interesting to compare the
result of the partial wave analysis with the one predicted by the
OPE model.

In this paper, we present the new data  on  the elastic and the
$pp\to pn\pi^+$ reactions measured at the proton momentum 1683
MeV/c. We compare the data with the OPE model calculations and
determine contributions of the different partial waves from the
combined partial wave analysis of the present data and the data
measured earlier.

\section{Experiment}
\label{sec:1}

The description of the experiment performed at the PNPI 1 GeV
synchrocyclotron was given in details in our previous work
\cite{Sarantsev:2004pe}. The proton beam was formed by three bending
magnets and by eight quadrupole lenses. The mean incident proton
momentum value was inspected by the kinematics of the elastic
scattering events. The accuracy of the incident momentum value and
momentum spread was about 0.5 MeV/c and 7 MeV/c (FWHM)
correspondingly with a perfect Gaussian distribution. A total of
$8\times10^4$ stereoframes were obtained. The frames were double
scanned to search for events due to an interaction of the incident
beam. The double scanning efficiency was determined to be 99.95$\%$.
Approximately 7$\times10^3$ two-prong events were used for the
subsequent analysis.

  The 2-prong events selected in the fiducial volume of the hydrogen
bubble chamber were geometrically reconstructed and kinematically fitted
to the following reaction hypotheses:
\begin{eqnarray}
  p+p &\to& p+p,                \\
  p+p &\to& p+n+\pi^+,          \\
  p+p &\to& p+p+\pi^0,          \\
  p+p &\to& d+\pi^+,            \\
  p+p &\to& d+\pi^++\pi^0.
\end{eqnarray}

 The  events identification procedure was also described in details in
\cite{Ermakov:2011at}. Thus, we list only the most severe criteria
here:
\begin{enumerate}
\item  Events with the confidence level larger than
$1\%$ were accepted.
\item  Events with only one acceptable hypothesis were
identified as belonging to this hypothesis.
\item If several versions revealed a good $\chi^2$ value, we
used the visual estimation of the bubble density of the track to
distinguish between proton (deuteron) and pion.
\end{enumerate}

The total number of the 2-prong events which had not pas\-sed the
reconstruction and fitting procedures was counted to be less than
$10\%$. These unidentified events were apportioned to the fraction
of the fitted hypotheses of the accepted events and were used only
for the total cross section calculations.
\begin{table}
\begin{center}
\caption{\label{tot_cs} Numbers of events and the total cross
sections at the beam momentum 1683 MeV/c. The total elastic cross
section was obtained by the interpolation of the differential cross
section by the Legendre polynomials. The errors include the
statistical errors and millibarn-equivalent ones. }

\begin{tabular}{|l||rc|}
\hline
$pp\to$ & events & $\sigma$  mb \\
\hline
elastic       & 2772 &  23.96   $\pm$ 0.57   \\
$pn\pi^+$     & 2564 &  18.97   $\pm$ 0.57   \\
$d\pi^+$      &   57 &   0.42   $\pm$ 0.05  \\
$d\pi^+\pi^0$ &    7 &   0.05   $\pm$ 0.02 \\
\hline
\end{tabular}
\end{center}
\end{table}

The standard bubble chamber procedure \cite{Andreev:1988fj} was used
to obtain absolute cross sections for the elastic and pion
production reactions. The precision in the determination of the
millibarn-equivalent was found to be $2\%$. The cross section values
for the inelastic processes together with statistics are listed in
Table~\ref{tot_cs}.
 Let us remind that data on the $pp\to pp\pi^0$ reaction at the same
energy were published earlier \cite{Sarantsev:2004pe}.

\begin{figure}[ht]
\centerline{\epsfig{file=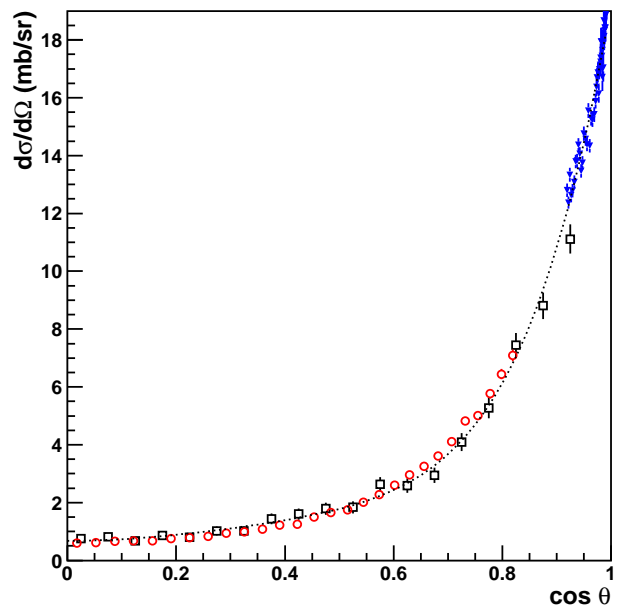,width=0.45\textwidth}}
\caption{\label{elastic} (Color online) Elastic differential cross
section. The dotted curve is result of the Legendre polynomial fit
of our data (open squares) restricted by the interval 0$\leq \cos\
\theta \leq$0.95 and the data from \cite{Dobrovolsky:1987wz} (blue
triangles). The open red circles show the measurements of the EDDA
experiment \cite{Albers:2004iw} taken at the incident momentum
1687.5 MeV/c.}
\end{figure}

The differential cross section for the elastic $pp$ scattering
measured in the present experiment is shown in Fig.~\ref{elastic} as
open squares with statistical errors. The value of the differential
cross section for the very forward angle bin is not shown in Fig.1
due a notable loss of events with a slow final proton. If the proton
momentum is less than 80 MeV/c the recoil paths is too short to be
seen in the bubble chamber. The events with the proton momentum less
than 200 MeV/c also might be missing during scanning. Since we do
not know the real amendment for these angles we excluded the last
forward point and only show the angles where the proton momentum is
above 200 MeV/c. In Fig.~\ref{elastic} we compare our elastic
differential cross section with the data from the EDDA experiment
\cite{Albers:2004iw} taken at the incident momentum 1687.5 MeV/c
(open red circles). One can see that there is a good agreement
between our points and the EDDA data, that supports the correctness
of our definition of the millibarn-equivalent.

To obtain the total elastic cross section we applied the following
procedure. We fitted the differential cross section with a sum of even
order Legendre polynomials $A_n\,P_n(\cos \theta)$ $n=0,2,4,\ldots$.
By examining the flatness of the behavior of the fit with decrease of the
fitted angular range we determined the range 0$\leq \cos\theta
<$0.95 as unbiased one.
 For finding the total elastic cross section we included
in the fit above $\cos\theta=0.95$ the data from
\cite{Dobrovolsky:1987wz} at the incident momentum 1685.7 MeV/c
which provide an important constraint for
high order polynomials. The result of the fit is shown in
Fig.~\ref{elastic} as the dotted curve. The total elastic cross
section calculated as $2\pi A_0$ was found to be $23.96 \pm 0.57$ mb
which is close to the value given in \cite{Shimizu:1982dx}.

\section{The $pp\to pn\pi^+$ reaction and a comparison with
the OPE model}

The OPE model \cite{Suslenko:1984qp} describes the single pion
production reactions by the four pole diagrams with the $\pi^{0}$ or
$\pi^{+}$ exchanges (we should like to express the deep appreciation
to the authors \cite{Suslenko:1984qp} for the accordance of their
program code). In this model the intermediate state of the $\pi
N$-scattering amplitude confines itself to the $P_{33}$ wave only,
assuming the leading role of the $\Delta_{33}$-resonance.

\begin{figure}[ht]
\centerline{\epsfig{file=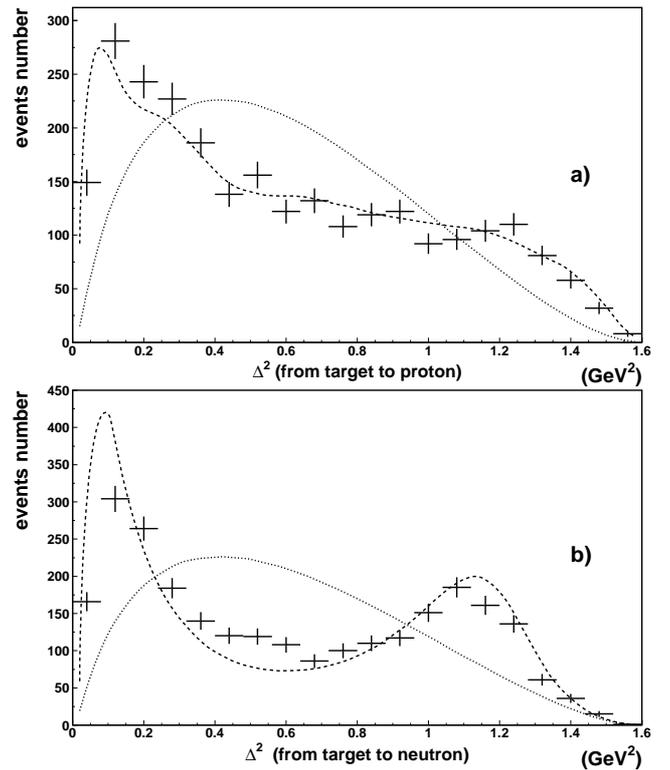,width=0.47\textwidth}}
\caption{\label{d2pip} Four-momentum transfer $\Delta^2$
distribution for the $pp \to pn\pi^+$ reaction: a) for the transfer
to the final proton and b) to neutron.  The dashed curves are the
OPE calculations and the dotted curves show the shape of the phase
volume.}
\end{figure}

 Fig.~\ref{d2pip} shows the distributions over the momentum transfer
squared, $\Delta^2 =-(p_t-p_f)^2$, where $p_t$ is the four-momen\-tum
of the target proton and $p_f$ is the four-momen\-tum of the final
proton or neutron in the $pp \to pn\pi^+$  reaction correspondingly.
The OPE model calculations normalized to the total number of the
experimental events is shown by dashed lines and the shape of the
phase volume is shown by dotted lines. One can see that the OPE
model describes qualitatively well the $\Delta^2$ distributions for
this reaction.

Fig. 3 presents c.m.s. angular distributions, effective two-particle
mass spectra of the final particles and angular distributions in the
helicity frame. We would like to point out that the c.m.s. angular
distributions are symmetrical ones which is a critical test for the
correctness of our event selection.

\begin{figure}[ht]
\centerline{\epsfig{file=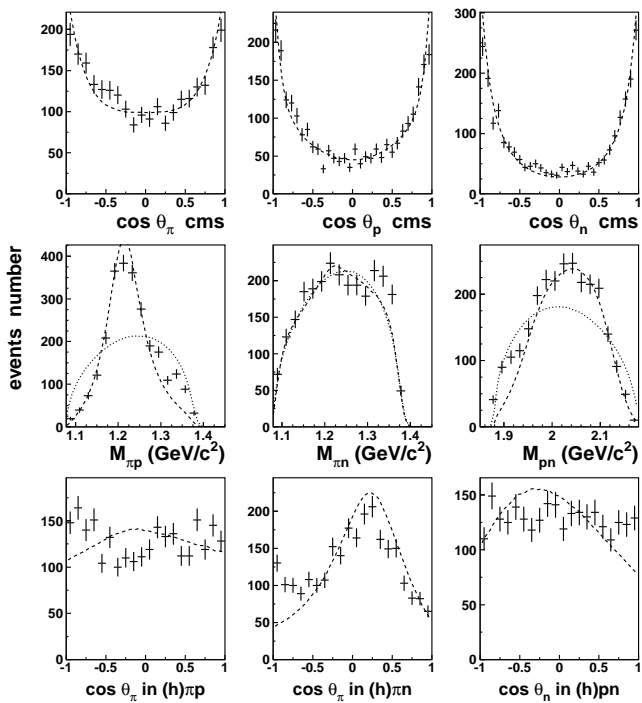,width=0.47\textwidth}}
\caption{\label{pip21ex} The $pp\to pn\pi^+$ data (the crosses with
statistical errors): angular distributions of the final particles in
the c.m.s. of the reaction (upper line), the effective two-particle
mass spectra (middle line) and angular distributions in the helicity
systems. The dashed curves show the OPE calculations and dotted
curves show the shape of the phase volume.}
\end{figure}

It is seen that the OPE model calculations normalized to the total
number of the experimental events reproduce the particle angular
distributions in the c.m.s. of the reaction and the two body mass
spectra fairly well. However the angular distributions in the
helicity systems show notable deviations from the experimental
points.

\section{Formalism of the partial wave analysis}

The partial wave analysis was performed in the framework of the
event-by-event maximum likelihood method. The formalism is given in
details in \cite{Anisovich:2007zz,Anisovich:2006bc} and based on the
spin-orbital momentum decomposition of the initial and final partial
wave amplitudes. Therefore it is natural to use the spectroscopic
notation $^{2S+1}L_J$ for two particle partial waves with the
intrinsic spin $S$, the orbital momentum $L$ and the total spin $J$.
Here and below we use $S,L,J$ for the description of the initial
$NN$ system, $S_2,L_2,J_2$ for the system of two final particles and
$S',L',J'\!=\!J$ for the system formed by the two-final particle
system and the spectator.

The total amplitude can be written as a sum of partial wave
amplitudes as follows \cite{Anisovich:2007zz,Anisovich:2006bc}:
\be
A=\sum\limits_\alpha A^\alpha_{tr}(s) Q^{in}_{\mu_1\ldots\mu_J}(S
LJ)A_{2body}^{S_2,L_2,J_2}(s_i)\times
\nonumber\\
Q^{fin}_{\mu_1\ldots\mu_J}(i,S_2L_2J_2S'L'J)\ ,
\ee
where $Q(S,L,J)$ are operators which describe the system of the
initial nucleons, $A^\alpha_{tr}$ is the transition amplitude and
$A_{2body}^{S_2,L_2,J_2}$ describes rescattering processes in the
intermediate two-particle channel. The multi-index $\alpha$ includes
all quantum numbers for the description of the definite partial
wave, $s$ is the invariant energy of the initial $NN$ system squared
and $s_i$ is the invariant energy squared of the two-particle
system.

To suppress contributions of amplitudes at high relative momenta we
introduced the Blatt-Weisskopf form factors. Thus the energy
dependent part of the partial wave amplitudes with production of a
resonance, for example,  in the two-particle system 12 (e.g. $\pi
p$) and the spectator particle 3 ($n$) has the form:
\be
A=\frac{A^\alpha_{tr}
A_{2body}^{S_2,L_2,J_2}(s_{12})q^{L}k_3^{L^{\prime}}}
{\sqrt{F(q^2,L,R)F(k_3^2,L^{\prime},r_3)}},
\label{atot}
\ee
where  $q$ is the momentum of the incident proton and $k_3$ is the
momentum of the spectator particle, both calculated in c.m.s. of the
reaction. The explicit form of the Blatt-Weisskopf form factors
$F(k^2,L,r)$ can be found, for example, in \cite{Anisovich:2004zz}.
One should expect that the effective radius of the initial
proton-proton system $R$ should vary between $1\div4$ fm. However,
due to a relatively large distance from the $pp$ threshold it is
hard to expect that this value can be determined with a good
accuracy in the present analysis. Indeed we did not observe any
sensitivity to this parameter and fixed it at 1.2 fm. A very similar
result was observed for $r_3$. So for our final fits we also fixed
this parameter at 1.2 fm.

The combined analysis of the data sets at different energies allows
us to extract the energy dependence of the partial waves which is
assumed to be a smooth function in this energy interval. This energy
dependence was introduced in the following form:
\be
A^\alpha_{tr}(s)=\frac{a^\alpha_{1}+a^\alpha_{3}\sqrt{s}}{s-a^
\alpha_{4}}e^{ia^\alpha_2}\,,
\ee
where $a^\alpha_i$ are real parameters. The $a^\alpha_4$ parameters
define poles located in the region of left-hand side singularities
of the partial wave amplitudes. Such poles are usually a good
approximation of the left-hand side cuts defined by the boson
exchange diagrams. The phases $a^\alpha_2$ are defined by
contributions from logarithmic singularities connected with three
body rescattering in the final state.

For the description of the energy dependence in the $\pi N$ system
we introduce two resonances: $\Delta(1232)\frac 32^+$ and Roper
$N(1440)\frac 12^+$. The corresponding amplitudes are parameterized
as follows:
\be
A^{S_2,L_2,J_2}_{2body}(s_{12})&=&\frac{k^{L_2}_{12}}
{\sqrt{F(k^2_{12},L_2,r_{12})}}
\frac{1}{M_R^2-s_{12}-M_R\Gamma},\nonumber\\
\Gamma&=&\Gamma_R \frac{M_R\,k_{12}^{2L_2+1}\,
F(k^2_R,L_2,r_{12})}{\sqrt{s_{12}}\,k^{2L_2+1}_R\,F(k^2_{12},L_2,r_{12})}\,.
\ee
Here $s_{12}$ is the invariant energy squared in the channel 12,
$k_{12}$ is the relative momentum of the particles 1 and 2 in their
rest frame  and $r_{12}$ is the effective radius.

 For $\Delta(1232)$, we use $M_R$ and $\Gamma_R$ taken from PDG
\cite{Beringer:1900zz} with $r_{12}=0.8$ fm. The Roper state was
parameterized using couplings found in the analysis
\cite{Anisovich:2011fc} where the decay couplings of this state into
the $\pi N$, $\Delta\pi$ and $N(\pi\pi)_{S-wave}$ channels were
determined.

For the description of the final $NN$ interaction we use the
following parameterization:
\begin{equation}
A^{S_2,L_2,J_2}_{2body}(s_{23})=\frac{\sqrt{s_{23}}}{1\!-\!\frac 12
r_{23}^\beta k_{23}^{2}a^\beta\!+\!ik_{23}a^\beta
\frac{k_{23}^{2L_2}}{F(k_{23},r_{23}^\beta,L_2)}}\,.
\label{a_2b}
\end{equation}
For the $S$-waves it coincides with the scattering-length
approximation formula suggested in \cite{watson,migdal}. Thus the
parameter $a^\beta$ can be considered as the $NN$-scattering length
and $r^\beta$ is the effective range of the $NN$ system.

\section{Partial wave analysis results and discussion}

\begin{figure}[ht]
\centerline{\epsfig{file=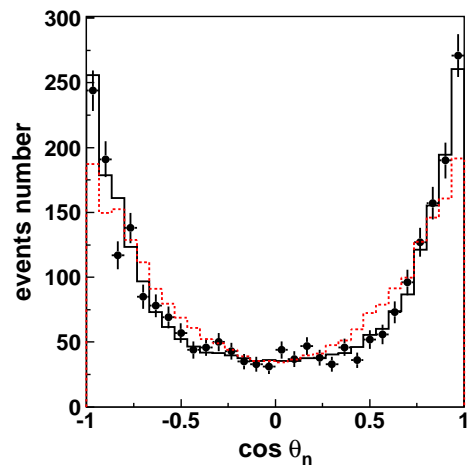,width=0.35\textwidth}}
\caption{\label{low_high} (Color online) The neutron angular
distribution calculated in the c.m.s. of the $pp\to pn\pi^+$
reaction at 1683 MeV/c. The data are shown by black circles with the
statistical errors. The solid (black) histogram shows the prediction
from the solution \cite{Ermakov:2014aj} with including partial waves
up to $L=5$ and the dotted (red) histogram shows the prediction from
the solution with $L$ up to 3.}
\end{figure}

\begin{figure*}[ht]
\centerline{\epsfig{file=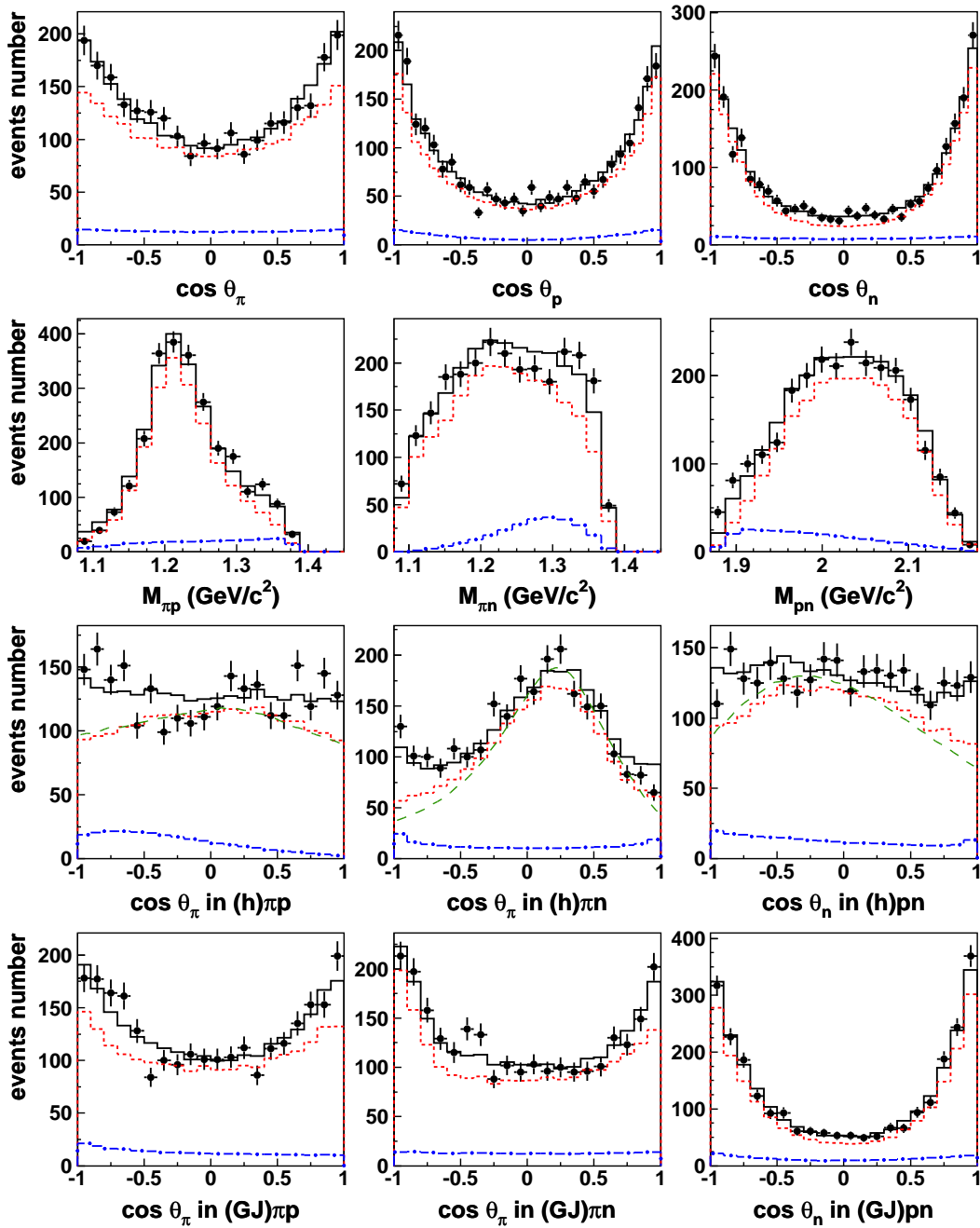,width=0.78\textwidth}}
\caption{\label{pwa21ex} (Color online) The $pp\to pn\pi^+$ data
taken at the proton momentum 1683 MeV/c with the statistical errors
only. First line: the angular distributions of the final particles
in the c.m.s. of the reaction. Second line: the effective
two-particle mass spectra. Third line: the angular distributions of
the final particles in the helicity frame. Fourth line: the angular
distributions of the final particles in the Gottfried-Jackson frame.
The solid (black) histograms show the result of our partial wave
analysis; the dotted (red) and dot-dashed (blue) histograms show the
contributions from the production of the  $\Delta(1232)$ and
$N(1440)$ intermediate states. The dashed (green) curves in the
helicity frame show the normalized distributions from the OPE
model.}
 \end{figure*}

We have performed the analysis of the new data starting from our
solution obtained in \cite{Ermakov:2014aj}. This solution was
restricted by the partial waves with the total spin $J$ up to 2 and
the orbital momentum $L$ up to 3. This solution produced an
acceptable description of the lower energy data but has notable
problems in the description of the new data set. For example, the
$\chi^2$ for the normalized angular distribution of the neutron in
the c.m.s. of the reaction is equal to 4.49. The solution fails to
describe the extreme angles which are mostly sensitive to partial
waves with high orbital momentum. Indeed, the solution with $L\leq
5$ and $J\leq 4$ found in \cite{Ermakov:2014aj} (but only used for
the error estimation in that paper) predicts the $\chi^2$ to be
1.23. The description of the data with these two solutions is shown
in Fig.~\ref{low_high}. This provides a strong argument for the
presence of higher partial waves at studied energy.

Although the solution with $L\leq 5$  produced a rather good
description of the normalized differential cross section, the total
cross section predicted by both solutions appeared to be about 10\%
lower than that given by the data. Therefore we used the last
solution as a starting point and performed the combined fit of the
present data together with the $pp\to pp \pi^{0}$ data measured
earlier \cite{Andreev:1988fj,Sarantsev:2004pe,ElSamad:2006wy} and
the $pp\to pn \pi^{+}$ data taken at 1628 and 1581 MeV/c
\cite{Ermakov:2011at,Ermakov:2014aj}.

On this way we were able to reproduce both the differential and the
total cross sections for all fitted data with a good accuracy. It's
worth to note that there is no problem to describe simultaneously
the total cross sections for the $pp\to pn \pi^{+}$ and $pp\to pp
\pi^{0}$ reactions in this approach.

The result of the partial wave analysis is shown in
Fig.~\ref{pwa21ex}: the histograms correspond to the Monte Carlo
events weighted by the differential cross section calculated from
the fit parameters. The $\chi^2$ for the distributions shown in this
picture is varied from 0.65 (for the pion angular distribution in
the c.m.s. of the reaction) to 2.6 (for the $\pi p$ invariant mass).
We would like to remind that we use the event-by-event maximum
likelihood analysis and do not fit directly these distributions.

The partial wave analysis (PWA) reproduces rather well the angular
distributions in the helicity system which have systematic
deviations in the OPE model. The OPE predictions normalized to the
contribution from the $\Delta(1232)$ production calculated from the
PWA solution are shown in Fig.~\ref{pwa21ex} with the dashed lines.
It is seen that $\Delta(1232)$ production from the partial wave
analysis and from the OPE model corresponds well each to another.
This confirms that $\Delta(1232)$ is produced by the one pion
exchange mechanism and the deviation of the data from the OPE model
is due to production of the Roper state.

The present combined analysis found the contributions from the
leading initial partial waves to be in a qualitative agreement with
the prediction from the solution reported in \cite{Ermakov:2014aj}.
However we observe changes for the contributions of the initial
partial waves $^1D_2$ and $^3F_2$ which are notably increased after
the fit of the new data. As concern the partial waves with the total
spin $J=4$ we found a sizeable contribution from $^3F_4$.

For all initial partial waves the contribution of channels with the
$\Delta(1232)$ production varies from 65 to 100$\%$ and only for the
$^3P_0$ wave it was found to be rather small one: 12$\%$. The Roper
resonance is produced mostly (in the decreasing order of
contributions) from the $^3P_2$, $^3P_0$, $^3P_1$ states and by one
order smaller from $^1S_0$. We found a notable contribution for the
decay of the initial $^3P_2$ state into the ($pn$) subsystem $^1P_1$
with isospin $I=0$.

\begin{figure}[ht]
\centerline{\epsfig{file=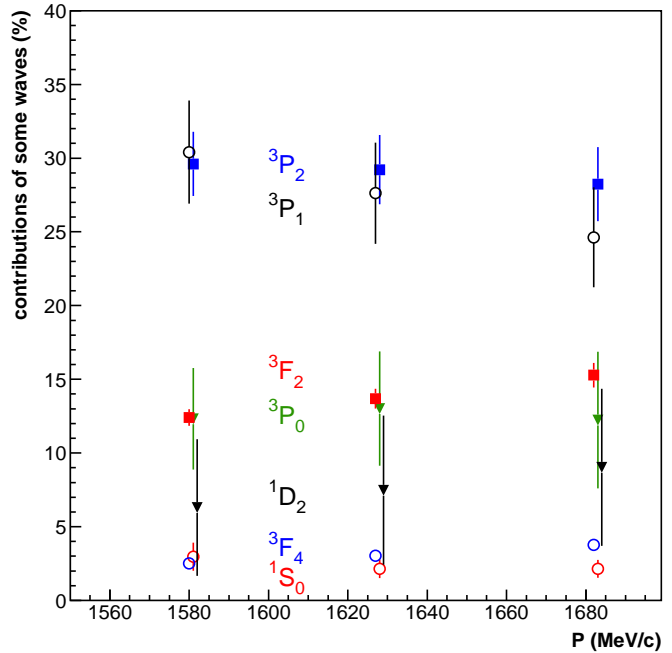,width=0.48\textwidth}}
\caption{\label{contr} (Color online) Contributions (the percentages)
of most important waves in the $pp \to pn\pi^+$ reaction.}
\end{figure}

To study the stability of the solution we added to the fit partial
waves with the total spin $J$ up to 5 decaying into $\Delta(1232)N$.
The obtained solution demonstrated some reduction of the
contribution from the $^3P_0$ initial state and increasing  the
contributions from the $^3F_2$ state. Taking into account these
ambiguities we have performed an error analysis of the initial state
contributions to the single pion production cross sections. For the
$pp\to pn\pi^+$ reaction these contributions are shown for three
incident momenta in Fig. ~\ref{contr}.

 It is necessary to mention that the
present combined analysis defines contributions of the partial waves
with smaller errors than it was found in \cite{Ermakov:2014aj}: the
unstable contributions from the high spin amplitudes are fixed with
the present data.

The obtained solution is well compatible with the data of HADES
collaboration on the single pion production at the energy 1.25 GeV
\cite{Agakishiev:2015piv}: including these data in the combined fit
does not change the main results of the analysis and contribution of
the partial waves to the HADES data was found to be in the errors
given in \cite{Agakishiev:2015piv}. The combined analysis of our and
HADES data should be a subject of the future joint partial wave
analysis.

\section{Conclusion}

The new data on the elastic and $pp\to pn\pi^+$ reactions taken at
the incident proton momentum 1683 MeV/c are reported. Including
these inelastic data in the combined partial wave analysis of the
single pion production reactions leads to a better error analysis
and  therefore to a more precise definition of the partial wave
contributions to the $pp\to pn\pi^+$ reaction.  We observe some
changes and in specific transition amplitudes compared to the
predictions from the solution \cite{Ermakov:2014aj}.

As noted earlier in Ref.~\cite{Andreev:1988fj}, although the OPE
model provides a qualitative description of most differential
distributions, it fails to describe simultaneously the total cross
section of the $pp\to pp\pi^0$ and $pp\to pn\pi^+$ reactions in the
investigated energy region. However our partial wave analysis
confirms the dominant role of the $\Delta(1232)$ production defined
by the OPE exchange mechanism. The main source of the discrepancy
between OPE and experimental data is due to contribution of other
intermediate states, in particular the Roper resonance.

The all analyzed data sets can be downloaded from the Bonn-Gatchina
data base \cite{webpage} as 4-vectors and directly used in the
partial wave analysis by other groups. We would like to remind that
although we supply a Monte Carlo sample in our web page one can use
a standard sample of $4\pi$ generated events: the bubble chamber
events have the efficiency which is close to 100\%.

\begin{acknowledgement}
We would like to express our deep gratitude to the bubble
chamber staff as well as to laboratory assistants, which toiled at
the film scanning and measuring. The work of V.A.Nikonov and
A.V.Sarantsev is supported by the RNF grant 16-12-10267.
 \end{acknowledgement}

\end{document}